\documentclass{article}

\usepackage{arxiv}

\usepackage[utf8]{inputenc} 
\usepackage[T1]{fontenc}    
\usepackage{hyperref}       
\usepackage{url}            
\usepackage{booktabs}       
\usepackage{amsfonts}       
\usepackage{nicefrac}       
\usepackage{microtype}      
\usepackage{lipsum}		
\usepackage{graphicx}
\usepackage{doi}
\usepackage{algorithm2e}
\usepackage{array}
\usepackage{comment}
\usepackage[numbers]{natbib}
\usepackage{amsmath}

\title{ChainGuards: Verification of Sensed Data\\using Permissioned Blockchain Technology}

\author{
    \href{https://orcid.org/0009-0005-5543-3588}
    {\includegraphics[scale=0.06]{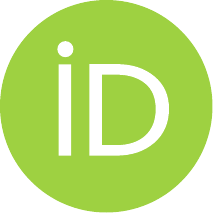}
    \hspace{1mm} Sara Aguincha}\\
	INESC-ID, Instituto Superior Técnico,\\Universidade de Lisboa, Portugal.\\
	\texttt{sara.aguincha@tecnico.ulisboa.pt} \\
	\And
    \href{https://orcid.org/0009-0003-2081-7712}
    {\includegraphics[scale=0.06]{orcid.pdf}
    \hspace{1mm} Emanuel Nunes} \\
    Sensefinity.\\
    Lisboa, Portugal.\\
	\texttt{emanuel.nunes@sensefinity.com} \\
    \And
	\href{https://orcid.org/0000-0003-0972-4171}
    {\includegraphics[scale=0.06]{orcid.pdf}
    \hspace{1mm} Samih Eisa} \\
    INESC-ID, Instituto Superior Técnico,\\Universidade de Lisboa, Portugal.\\
	\texttt{samih.eisa@inesc-id.pt} \\
    \And
	\href{https://orcid.org/0000-0003-2872-7300}
    {\includegraphics[scale=0.06]{orcid.pdf}
    \hspace{1mm} Miguel L. Pardal} \\
    INESC-ID, Instituto Superior Técnico,\\Universidade de Lisboa, Portugal.\\
	\texttt{miguel.pardal@tecnico.ulisboa.pt} \\
}

\renewcommand{\shorttitle}{ChainGuards}

\hypersetup{
pdftitle={ChainGuards: Verification of Sensed Data using Permissioned Blockchain Technology},
pdfsubject={Blockchain, Internet of Things, Data Integrity},
pdfauthor={Sara Aguincha, Emanuel Nunes, Samih Eisa, Miguel L. Pardal},
pdfkeywords={Sensors, Sensed Data, Safety, Internet of Things, Supply Chain},
}

\begin{document}
\maketitle

\begin{abstract}
Sensor technologies have evolved to a point where it is now practical to monitor products along the supply chain. The collected data can be stored in a decentralized way using blockchain technology. However, ensuring the reliability of the sensed data is a critical challenge. In other words, we need to trust the data that we write to the blockchain.

In this work, we propose \emph{ChainGuards}, a decentralized system that uses product-specific rules to verify data collected across the supply chain, with particular focus on sensor-derived information, issuing warnings and triggering audits when anomalies are detected. We evaluated \emph{ChainGuards} using data from a real cherry supply chain deployment. The result shows that the implemented solution provides reliable verification of supply chain data with low performance overhead, able to correctly detect data discrepancies and inconsistencies.
    
\end{abstract}


\section{Introduction}
\label{sec:introduction}
A supply chain is composed of a network of businesses and individuals, commonly referred to as trading partners, that collaboratively produce, transport, and distribute goods such as pharmaceuticals, electronics and food. However, ensuring the safety, authenticity and quality of products across these complex networks remains a major challenge. Recent scandals involving unsafe pharmaceutical ingredients \cite{dyer2018johnson}, defective aircraft components \cite{kotze2023airplane} and mislabeled meat products \cite{premanandh2013horse} highlight persistent weaknesses in supply-chain oversight. In particular, maintaining proper storage and transport conditions is important for preserving product quality and safety. Existing food safety regulations and certification schemes often rely on sporadic audits and trust in labeling practices, which may be incomplete, delayed or inaccurate.

The growing affordability and availability of Internet of Things (IoT) devices has made continuous monitoring of supply chains increasingly feasible. Sensors can now track environmental conditions and product movement from production to retail. However, while such technologies improve visibility, they raise a fundamental question: \textit{how can we ensure that the data collected by sensors is reliable?}

Sensor data may be affected by hardware faults, noise, calibration drift, communication failures or deliberate tampering. As a result, trustworthy supply-chain information requires mechanisms not only to authenticate sensors but also to validate their reported data and detect inconsistencies or anomalies. When violations occur, stakeholders should be alerted to intervene before quality or safety is compromised. A central challenge lies in determining who should define and enforce the rules used to verify the data. Ideally, these rules should be established collaboratively by a consortium of domain experts representing all stages of the supply chain. For example, in agricultural context, food safety experts may specify acceptable temperature and humidity ranges, while logistics professionals define constraints on transport duration and handling. The verification criteria can also be shaped by industry best practices and legal requirements.

When the data is initially collected, it reflects the perspective of a single participant and cannot yet be considered an objective fact by all trading partners. Verification is therefore required before such data can be trusted collectively. Relying on a centralized authority for this process would reintroduce single points of failure and undermine decentralization. Blockchain technology \cite{nakamoto2008bitcoin} offers a foundation for decentralized data verification. Distributed ledgers enable secure data sharing among supply-chain partners while supporting the shared execution of verification logic. By embedding validation directly within the blockchain, verification outcomes can be made transparent, auditable and tamper-resistant.

In this work, we propose \emph{ChainGuards}, a chaincode framework for permissioned blockchains that can be used to define and enforce product-specific verification rules on supply chain data, with emphasis on sensor-derived information. Data is initially recorded on the blockchain as claims made by individual stakeholders (e.g., \textit{“the supplier claims to have shipped 300 kg of product”}). As additional data becomes available, from sensors and information systems, \emph{ChainGuards} evaluates these claims against jointly defined rules. Once verified through decentralized execution, claims can be accepted as facts agreed upon by the consortium (e.g., \textit{“300 kg of product was shipped”}).

We evaluate \emph{ChainGuards} using data from FoodSteps project (Section~\ref{sec:FoodSteps}), which monitors real-world agricultural supply chains using temperature, humidity and location sensors. Our results show that \emph{ChainGuards} effectively detects faulty or inconsistent data while introducing minimal performance overhead, enabling trustworthy supply-chain monitoring without disrupting operational workflows.

\section{Background and Related Work}
\label{sec:bg-rw}

This section summarizes foundations and prior work relevant to \emph{ChainGuards}. We review sensor data integrity challenges in IoT systems, blockchain-based traceability with emphasis on permissioned ledgers, and existing approaches to sensor data verification. We also position \emph{ChainGuards} within the literature on business rules as an explicit and auditable way to express verification logic.

\subsection{Sensor Data Integrity}
The increasing deployment of IoT systems, from smart homes to large-scale infrastructures, has intensified concerns about the reliability of sensor-generated information. A useful distinction is between \emph{safety} and \emph{security}. Lyu et al.~\cite{lyu2019safety} describe \emph{physical safety} as the prevention of hazards such as fires or floods, and \emph{functional safety} as maintaining safe system behavior under accidental failures. \emph{Security}, in contrast, focuses on protecting systems against intentional threats (e.g., tampering or unauthorized access) to preserve confidentiality, integrity and availability.

Sensor data is inherently error-prone. Teh et al.~\cite{teh2020sensor} survey common data-quality issues including outliers, missing data, bias, drift, noise, constant-value failures, uncertainty and stuck-at-zero behavior. Such faults may result from device degradation, harsh environmental conditions, improper handling, or intermittent connectivity. Addressing these issues purely through higher-grade hardware or frequent retransmission is often impractical due to cost, energy constraints, and scalability limitations. Instead, systems typically rely on structured data processing pipelines, including \emph{preprocessing} (cleaning, integration, transformation, and reduction)~\cite{alasadi2017review}, \emph{error detection} to identify anomalies~\cite{xie2017anomalyDetection}, and subsequent analysis. When redundant devices are available, \emph{multi-sensor fusion} can further improve robustness by combining correlated readings into a more reliable estimate~\cite{tsanousa2022review}.


\subsection{Blockchain for Traceability}
Supply-chain monitoring often relies on data shared across organizational boundaries, where manual supervision or centralized validation can introduce delays, human error and manipulation risks. Blockchains, such as Bitcoin introduced by Nakamoto~\cite{nakamoto2008bitcoin}, provides a tamper-evident ledger in which transactions are grouped into blocks cryptographically linked by hashes. This structure makes retrospective modification difficult, while digital signatures authenticate transaction authorship and support non-repudiation.

In enterprise settings, permissioned blockchains are often preferred over public ledgers. Hyperledger Fabric~\cite{androulaki2018hyperledger} is a modular, permissioned framework that supports authenticated participation, configurable consensus, and fine-grained data sharing. Its architecture includes \emph{peers} that validate transactions and maintain ledger state, \emph{ordering services} that sequence transactions into blocks, \emph{channels} that isolate ledgers among subsets of participants, and \emph{chaincode} that implements application logic. Endorsement policies define which organizations must approve a transaction before it can be committed, enabling shared governance across a consortium.

Blockchain can thus ensure the integrity of stored records and provide a shared execution environment for verification logic. However, immutability alone does not guarantee the correctness of ingested sensor data, reflecting the adage \textit{``garbage in, garbage out''}. This limitation motivates the development of mechanisms, such as \emph{ChainGuards}, which aim to enhance data reliability.

\subsection{Supply Chain Traceability}
Supply chains span production, transformation, and distribution among diverse entities. In the food sector, the European Union defines traceability as \textit{``the ability to trace and follow a food, feed, food-producing animal or substance \ldots\ through all stages of production, processing and distribution''}~\cite{official2002journal}. Interoperability is supported by GS1 identifiers and information-sharing standards, including EPCIS (Electronic Product Code Information Services)~\cite{gs1traceability}. The EPCIS represents traceability as an ordered sequence of events capturing \emph{Who}, \emph{What}, \emph{When}, \emph{Where}, \emph{Why}, and optionally \emph{How}, providing a widely adopted foundation for cross-organizational traceability.

A growing body of work explores blockchain to complement these standards, improving auditability and transparency in domains such as wine authentication~\cite{wine2017blockchain}, dairy~\cite{dairy2021blockchain}, baked goods~\cite{carasau2021blockchain}, and fisheries~\cite{fishery2022blockchain}. Similar approaches exist for textiles~\cite{textile2021blockchain}, pharmaceuticals~\cite{drugs2021blockchain,ppecovid2022blockchain}, and mechanical parts~\cite{alsadi2023trucert}, as well as certification-focused scenarios such as halal compliance~\cite{halal2022blockchain}. Collectively, these systems show that blockchain can strengthen the integrity and transparency of traceability records, yet they typically assume that sensor inputs are reliable or validated externally.

\subsection{Sensor Data Verification and Business Rules}
Existing approaches to sensor data verification can be broadly grouped into three categories. \emph{Rule-based verification} relies on thresholds or consistency checks and is attractive for its transparency and low computational cost, but it can miss subtle faults and is limited against adversarial behavior~\cite{cook2020smart}. \emph{Cross-source validation} improves fault tolerance by comparing redundant measurements, though many proposals operate within a single organization and implicitly assume honest reporting. \emph{Learning-based methods} use statistical or machine learning models to detect anomalies; while powerful, they often require extensive training data and can be opaque, making them harder to justify in compliance-critical environments~\cite{ahmed2023survey}.

\emph{ChainGuards} extends rule-based verification with both intra-organizational redundancy (MSoD) and inter-organizational corroboration (DSoD), and it integrates these checks into the blockchain layer to ensure results are shared, consistent, and auditable across stakeholders. This design aligns naturally with the notion of \emph{business rules}: explicit statements that constrain or guide organizational behavior. Muehlen et al.~\cite{zur2010modeling} characterize business rules as statements intended to influence actions and information in an organization, while Boyer and Mili~\cite{boyer2011agile} argue that such rules should be externalized from application code to improve transparency and adaptability. Acharya~\cite{acharya2012oracle} further emphasizes Event--Condition--Action (ECA) structures for event-driven decision automation. In \emph{ChainGuards}, verification policies adopt the same principle: rules are explicitly represented, configurable, and executed in response to supply-chain events, enabling consistent enforcement and producing an auditable trail of warnings and alerts.

Prior work establishes strong foundations for sensor anomaly handling and blockchain-based traceability, but leaves a gap in decentralized, stakeholder-governed verification of sensor data. \emph{ChainGuards} addresses this gap by combining policy-driven rules with cross-source corroboration and on-chain auditability.
Next, we introduce a system that provides context to demonstrate the use of \emph{ChainGuards}.

\section{FoodSteps}
\label{sec:FoodSteps}

In the agri-food supply chain, access to reliable and transparent product information is essential for informed food consumption. This includes data on origin and provenance, processing history, environmental conditions, and, where available, quality or sustainability indicators.

\begin{figure}[htp]
    \centering
    \includegraphics[width=0.99\columnwidth]{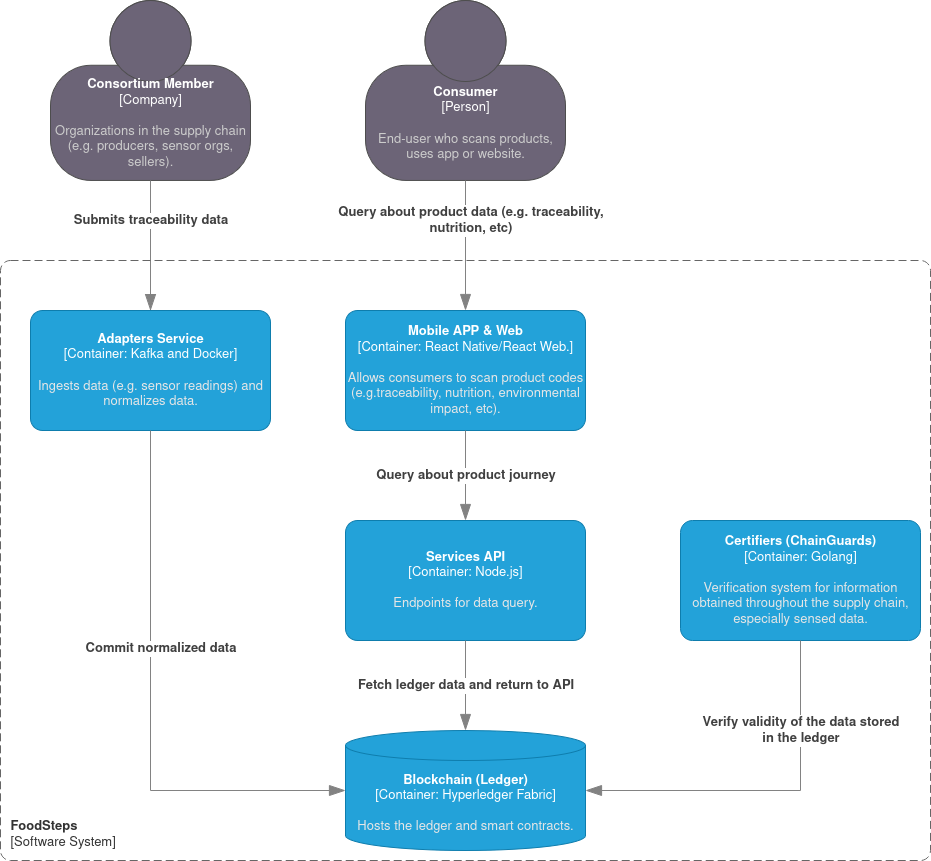}
    \caption{FoodSteps software architecture overview.}
    \label{fig:foodsteps}
\end{figure}

\textit{FoodSteps} is a blockchain-based food traceability platform that records and exposes each stage of a product’s journey from farm to fork. Its architecture is shown in Figure~\ref{fig:foodsteps} using C4 notation\footnote{Context, Containers, Components, and Code diagrams. \url{https://c4model.com/}}.

Consortium members comprise supply-chain stakeholders such as producers, processors, logistics providers, warehouses, and sensor operators. Whenever a stakeholder collects provenance-related data, such as sensor readings, transport events, or processing records, it is submitted to the Adapters Service. This service validates incoming data against a canonical schema, normalizes units, enriches records with metadata, and packages them into transactions that are committed to a Hyperledger Fabric ledger, ensuring immutability and auditability.

Consumers interact with FoodSteps through a web or mobile application. By scanning a QR code or entering a product identifier, users query the Services API, which retrieves verified provenance information from the blockchain. The response includes the product’s traceability history, contextual metadata, and verification outcomes, enabling transparent inspection of the product’s lifecycle.

At the core of FoodSteps is a permissioned Hyperledger Fabric network. Each organization operates a private channel for sensitive operational data, which remains confined until explicitly approved for sharing. When stakeholders agree to disclose specific data, for example, after a sale is completed, relevant records are migrated to a shared channel accessible only to authorized parties. This channel-based access control enables end-to-end traceability while preserving commercial confidentiality, as discussed in the work by Nunes et al.~\cite{Nunes25cc2c}.

\section{ChainGuards Design}
\label{sec:chainguards}

\emph{ChainGuards} is a decentralized verification framework designed to increase trust in supply-chain data, with particular focus on sensor-derived information. As sensors are increasingly used to monitor product safety, authenticity, and origin, ensuring reliability of their outputs becomes important. Faulty, missing, or manipulated sensor readings can undermine confidence in complex, multi-organizational supply chains. \emph{ChainGuards} addresses this challenge by enforcing product-specific verification rules collaboratively defined by supply-chain stakeholders. These rules detect and flag inconsistencies before data is accepted as reliable, while verified records are immutably stored on the blockchain to support audits, regulatory compliance, and operational decision-making.

\subsection{System Overview and Participants}
\label{sec:system-overview}

\emph{ChainGuards} is integrated into the \textit{FoodSteps} system as an on-chain certification component (represented on the right side of Figure~\ref{fig:foodsteps}), implemented as Fabric chaincode. It is triggered when approved data is written into a designated channel, where it applies configurable verification rules to sensor readings and business events. Fabric was selected for its modular architecture, fine-grained access control, and suitability for consortium-based governance. The blockchain acts not only as a shared ledger but also as an execution environment for verification logic, ensuring that validation is performed transparently and consistently across all participants. Three categories of entities interact with the system: 
\begin{itemize}
    \item Clients, such as consumers or business users, who query verified product information;
    \item Consortium members, including producers, processors, logistics providers, and retailers, who operate blockchain peers, define verification policies and submit data;
    \item Auditors, independent entities notified when verification results produce warnings or alerts requiring further investigation.
\end{itemize}

The design of \emph{ChainGuards} was informed by an expert interview with a food safety and logistics specialist from a major national retailer from Portugal, focusing on perishable goods such as cherries. The expert highlighted key concerns including reliable harvest date recording for shelf-life estimation, verification of certified origin despite lot mixing, accurate location tracking during transport, monitoring of temperature and humidity exposure, integration of quality test results and improved recall procedures.
These insights translated into system requirements centered on provenance validation, multi-source data verification, discrepancy detection and audit support. Table~\ref{tab:system-requirements} summarizes the consolidated requirements that guided rule selection and system architecture.

\begin{table}[ht]
\centering
\footnotesize 
\caption[Requirements]{Key system requirements derived from expert insights and design}
\label{tab:system-requirements}
\begin{tabular}{ | >{\raggedright\arraybackslash}m{4cm} |
                   >{\raggedright\arraybackslash}m{10.7cm} | }
 \hline
 \textbf{Requirement} & \textbf{Description} \\ [0.5ex] 
 \hline
 R1: Product Origin & Verify products come from certified regions and/or producers. \\ 
 \hline
 R2: Location Verification & Ensure shipments depart/arrive in expected areas; Ensure correctness in taken routes. \\
 \hline
 R3: Environmental Monitoring & Detect deviations in temperature and humidity during handling. \\ 
 \hline
 R4: Quality Tests & Store and verify test results (e.g., Brix, pesticide) for traceability. \\ 
 \hline
 R5: Recall Procedures & Enable backward traceability for rapid recalls. \\ 
 \hline
 R6: Multi-Device Reliability & Aggregate data from multiple sensors to reduce faults. \\ 
 \hline
 R7: Auditor Notification & Notify an auditor when verification finds anomalies. \\ 
 \hline
 R8: Discrepancy Detection & Compare stakeholder reports to find inconsistencies. \\ 
 \hline
\end{tabular}
\end{table}

\subsection{Core Concepts and Data Model}
\label{design-concepts}

\emph{ChainGuards} represents supply-chain data using a hierarchical object model that supports verification at multiple abstraction levels:

\begin{itemize}
    \item A \textit{Journey} captures the lifecycle of a product abstraction (e.g., batch, pallet, or retail package), maintaining parent–child relationships across transformations and aggregations;
    \item A \textit{Step} corresponds to a business phase such as transport, storage, or processing, serving as a verification checkpoint;
    \item A \textit{Point} stores granular sensor readings (e.g., temperature sample or GPS coordinates) associated with a Step.
\end{itemize}

All objects are stored on-chain using composite keys, enabling efficient queries and preserving traceability. To limit ledger growth, raw sensor data is stored as independent Points, allowing concurrent ingestion from multiple devices or organizations without conflicts. Further details will be provided in the implementation (Section~\ref{sec:implementation}).

\subsection{Data Sources and Verification Model}
To accommodate varying levels of redundancy and trust, \emph{ChainGuards} supports three complementary verification models:

\begin{itemize}
    \item Single Source of Data (SSoD): validates data from one device or system, providing baseline assurance in low-instrumentation scenarios;
    \item Multiple Sources of Data (MSoD): cross-validates redundant data within a single organization, improving fault tolerance through multisensor fusion;
    \item Diverse Sources of Data (DSoD): compares independent claims from multiple organizations, enabling decentralized verification and discrepancy detection.
\end{itemize}

In addition, \emph{ChainGuards} explicitly models handover events, which capture custody transfers between organizations and verify consistency between departure and arrival records, supporting provenance validation and recall processes.

\subsection{Rules and Policies}
Verification behavior is governed by on-chain policy files, stored as JSON documents jointly approved by consortium members and then ``transferred'' to chaincode. Each policy specifies the product type, applicable verification model, relevant business phases and rule parameters. The current prototype implements the following rules:

\begin{itemize}
    \item \emph{Threshold Rule}: for environmental and temporal constraints; It monitors environmental conditions such as temperature or humidity to ensure they remain within safe limits (e.g., 2–8 °C for cherries). This rule directly supports product quality and shelf-life management;
     \item \emph{Geofence Rule}: to ensure shipments remain within authorized geographic regions; It enables verification of certified origins;
    \item \emph{Backtrack Detection Rule}: to identify unauthorized route deviations, contributing to shipment integrity and reducing risks of tampering or delays;
   \item \emph{Handover Time Rule}: to validate temporal consistency between custody transfers; It validates time consistency between consecutive custody transfer events, ensuring traceability during inter-organizational exchanges;
   \item \emph{Shipment Timeout Rule}: to detect excessive or implausible transit durations. It checks that the total duration of a shipment remains within expected limits, helping to detect transport delays that could impact product freshness or safety.
\end{itemize}

All rules are modular and extensible, allowing adaptation to specific products, regulatory frameworks and regional standards. New rule types can be integrated as additional requirements emerge.

\section {ChainGuards Implementation}
\label{sec:implementation}
Following the design in the previous Section, this Section describes the implementation of \emph{ChainGuards}.  It outlines the technological stack, core object model and processing workflow, including data ingestion, verification triggering, sensor preprocessing, rule execution, and auditing. The goal is to clarify how design concepts are realized in a working decentralized system.

\subsection{Technological Stack}

\emph{ChainGuards} is implemented as chaincode running on a Hyperledger Fabric blockchain. The chaincode is written in the Go programming language and was initially developed and tested using the Hyperledger Fabric Visual Studio Code extension. Deployment and network orchestration are automated using the Fablo framework, with Docker providing isolated and reproducible execution environments.

To support evaluation, a custom toolkit was developed using Python and Bash to generate synthetic supply-chain events and sensor data. Events are represented in JSON format, while environmental and GPS readings are generated as CSV files. Geographic routes are visualized using Python-based scripts that produce interactive HTML maps.

An external auditor notification service was also implemented in Go. It subscribes to Fabric events and notifies auditors when warnings or alerts are generated. In the prototype, notifications are delivered through Discord webhooks, although the design can be extended to support other messaging mechanisms.

\subsection{Object Model}

\begin{figure}[htp]
    \centering
    \includegraphics[width=0.8\columnwidth]{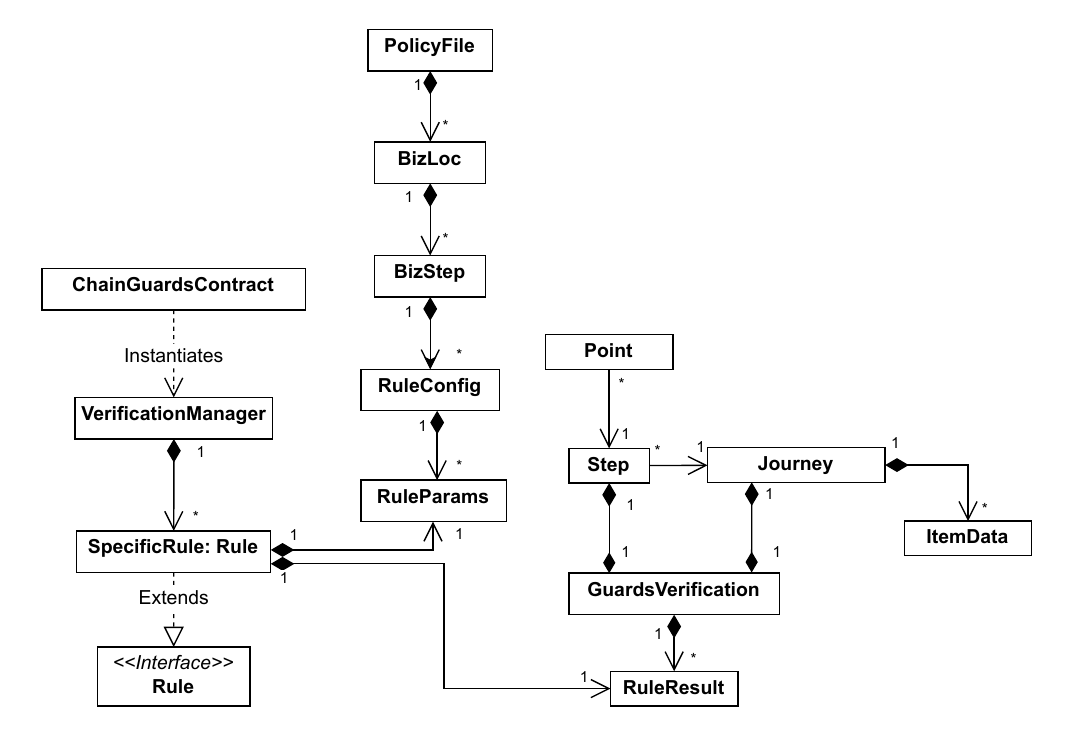}
    \caption{ChainGuards object model (UML)}
    \label{fig:chainGuardsArchitectureUML}
\end{figure}

\emph{ChainGuards} organizes supply-chain data through a hierarchical object model aligned with the conceptual design. Figure~\ref{fig:chainGuardsArchitectureUML} presents an overview of the main objects and their relationships.

\begin{figure}[htp]
    \centering
    \includegraphics[width=0.9\columnwidth]{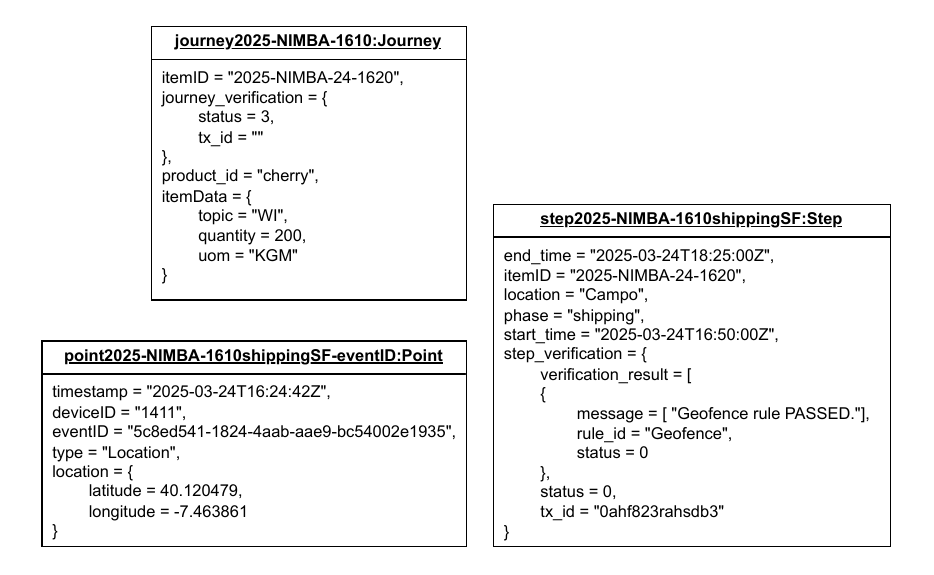}
    \caption{Example of Journey, Step, and Point objects}
    \label{fig:journey-step-point}
\end{figure}

The core concepts are \textit{Journey}, \textit{Step}, and \textit{Point}, representing progressively finer-grained views of a product’s lifecycle (as discussed in Section~\ref{sec:chainguards}). Figure~\ref{fig:journey-step-point} illustrates an example of these objects. Objects are stored in the Fabric world state using composite keys that encode their type and identifiers. This enables efficient partial queries, such as retrieving all Steps and Points associated with a given product. Raw sensor readings are stored independently as Points rather than embedded in Steps, allowing concurrent reporting from multiple devices or organizations without write conflicts. 
\emph{ChainGuards} adopts the EPCIS standard~\cite{gs1traceability} as the canonical format for representing supply-chain events. Journeys also store stakeholder-specific claims (e.g., quantity or quality attributes) using \textit{ItemData} objects, enabling later cross-party comparison. Verification logic is defined through on-chain \textit{Policy} objects. Each policy specifies the product type, verification mode (SSoD, MSoD, or DSoD), and the business locations and phases to which verification rules apply. Storing policies on-chain ensures shared visibility, immutability, and consistent enforcement across consortium members.
Verification is orchestrated by the \textit{VerificationManager}, which loads the relevant policy and executes the applicable rules for a given Step or Journey. Results are stored in a \textit{GuardsVerification} object, classifying outcomes as \textit{okay}, \textit{warning}, or \textit{alert}. This separation enables independent step-level verification and aggregation at the journey level.

\subsection{Canonical Formats}

\emph{ChainGuards} is invoked with a JSON payload containing a topic identifier and an EPCIS event. The topic identifies the data owner or source channel, which is required to detect discrepancies between stakeholder claims. These EPCIS events follow the standard's model for representing business actions and sensor observations, with limited adaptations to support project-specific requirements.

For sensor preprocessing, internal data structures group Points by sensor type and device. Each device is tracked with a dynamic reliability score, which is updated over time based on observed behavior. These scores influence weighted data fusion during verification, reducing the impact of faulty sensors. Policy files define verification rules and parameters, including thresholds, geofences, time windows, and severity levels. Verification outcomes are recorded with detailed rule-level results and linked to blockchain transaction identifiers to ensure full traceability.

\subsection{Processing Model}
\label{processing-model}

The processing model shows how data is ingested, processed, verified and recorded within \emph{ChainGuards}.

\subsubsection{Event Ingestion}

When a stakeholder shares EPCIS events to a channel monitored by \emph{ChainGuards}, the chaincode parses the event and extracts the relevant product identifiers. Based on this information, the system creates or updates the corresponding Journeys and Steps, while also recording the associated stakeholder claims. Sensor readings are handled independently and stored as Point objects, which can be retrieved and linked to their respective Steps.

Certain EPCIS events, such as aggregation or disaggregation, result in the creation of new Journeys (e.g., when pallets are split). In such cases, parent–child relationships are updated accordingly to preserve traceability.


Verification can be initiated either manually by authorized consortium members or automatically upon the occurrence of specific EPCIS events that signal the completion of a business phase (e.g., receiving events). However, due to the asynchronous nature of data submission and the possibility of delays, fully automated verification may be unreliable. Consequently, manual triggering serves as the primary mechanism to ensure data completeness, while event-based triggers provide partial automation.

\subsubsection{Sensor Data Preprocessing}
\label{preprocessing}

Before rule execution, sensor streams are filtered to reduce noise and false positives. For single source of data (SSoD), statistical outliers are removed using interquartile range (IQR) analysis for scalar data and speed-based checks for GPS data. The remaining signal is smoothed using a Kalman filter~\cite{kalman1960new}.

For multiples sources of data (MSoD), measurements originating from multiple devices are fused within the Kalman filter, with readings from the same time frame weighted using the Mahalanobis distance~\cite{mahalanobis1936generalised} at each update. If only a single device is available, the system falls back to the SSoD pipeline.

With diverse sources of data (DSoD), preprocessing remains identical to SSoD and MSoD. The distinction arises at the journey level verification, where data is compared across organizations.

\subsubsection{Verification and Recording}

After preprocessing, rules defined in the policy are executed. Most rules apply at the step level, while journey-level rules handle aggregation and hand-over consistency across organizations. In DSoD scenarios, claims from different stakeholders are compared to identify discrepancies.

Verification results are stored on-chain together with transaction identifiers. Warnings and alerts trigger auditor notifications through Fabric event listeners. All verification outcomes remain queryable and auditable, enabling retrospective analysis and re-verification if needed.

\subsection{Execution Example}

\begin{figure}[htp]
    \centering
    \includegraphics[width=0.99\columnwidth]{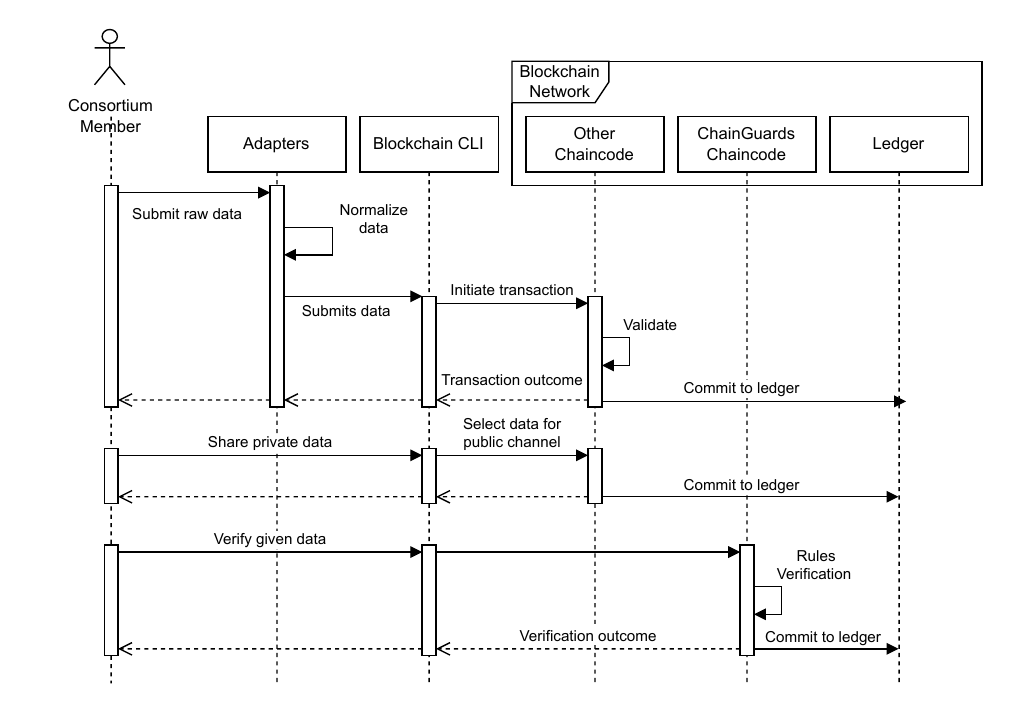}
    \caption{ChainGuards execution sequence}
    \label{fig:sequenceDiagramExecution}
\end{figure}

Figure~\ref{fig:sequenceDiagramExecution} illustrates a complete execution flow. Stakeholders first submit raw data to private channels. Once data becomes shareable, it is replicated to a public channel where \emph{ChainGuards} ingests and verifies it. Verification may be triggered manually or by key events, resulting in recorded verification outcomes and, if necessary, auditor notifications.

\section{Evaluation}
\label{sec:evaluation}

We evaluated \emph{ChainGuards} using a combination of simulated experiments and real data from a pilot deployment.
Simulations enabled controlled testing of all implemented rules under both valid and faulty conditions, while the pilot assessed the system’s behavior under realistic sensing and logistics constraints.

\begin{figure}[h]
    \centering
    \includegraphics[width=0.95\columnwidth]{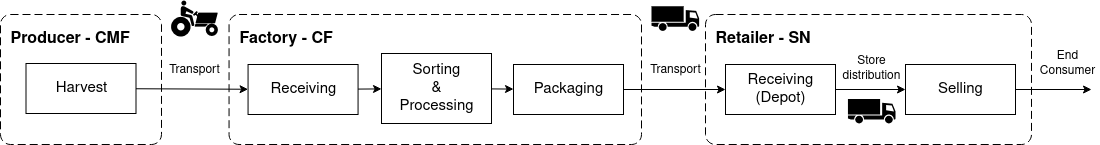}
    \caption{Supply chain of Fundão Cherry}
    \label{fig:usecase-supplychain}    
\end{figure}

The pilot is the Fundão Cherry supply chain, represented in Figure~\ref{fig:usecase-supplychain}, and detailed in Section~\ref{sec:pilot}.

\subsection{Rule Evaluation}

\emph{ChainGuards} enforces rules defined in configurable policy files, including: geofence, backtrack, threshold verifications, multi-sensor analysis (MSoD), and discrepancy detection (DSoD). 

A dedicated toolkit was developed to generate synthetic traceability events in EPCIS format, simulating geographic routes, timestamps and environmental readings. OSRM (Open Source Routing Machine)~\cite{luxen-vetter-2011} was used to generate plausible route coordinates and travel times.

\begin{figure}[htp]
    \centering
    \begin{minipage}[t]{0.425\textwidth}
        \centering
        \includegraphics[width=\linewidth]{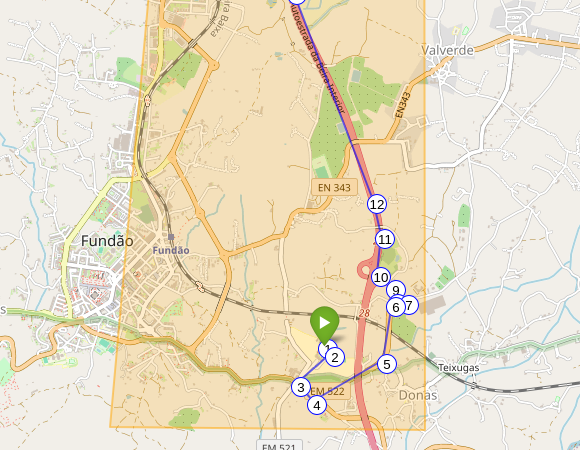}
        \caption{Valid geofence from Farm to Factory}
        \label{fig:valid-geofence}
    \end{minipage}
    \begin{minipage}[t]{0.425\textwidth}
        \centering
        \includegraphics[width=\linewidth]{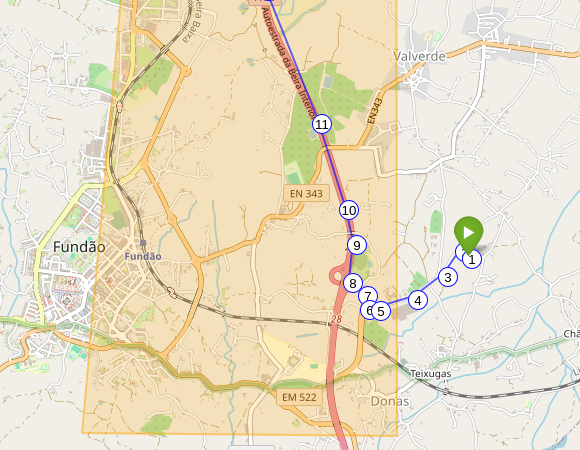}
        \caption{Invalid geofence from Farm to Factory}
        \label{fig:invalid-geofence}
    \end{minipage}\hfill    
\end{figure}

The \textit{geofence rule} ensures that the shipments remain within predefined geographical boundaries, which is particularly important for certified goods. We tested two simulated shipments scenarios along the main routes from a fruit field to the processing factory (Section ~\ref{sec:pilot}). Figure \ref{fig:valid-geofence} shows a shipment within a valid geofence, passing verification, while Figure \ref{fig:invalid-geofence} depicts a route from a nearby field violating the geofence rule, triggering an alert in the system.
The shipments also undergo radius checks to confirm departures and arrivals are within acceptable distances (according to the requirements stated in Section~\ref{sec:system-overview}). The valid shipment passes, whereas the invalid one fails due to an out-of-range start.

\begin{figure}[htp]
    \centering
    \includegraphics[scale=0.38]{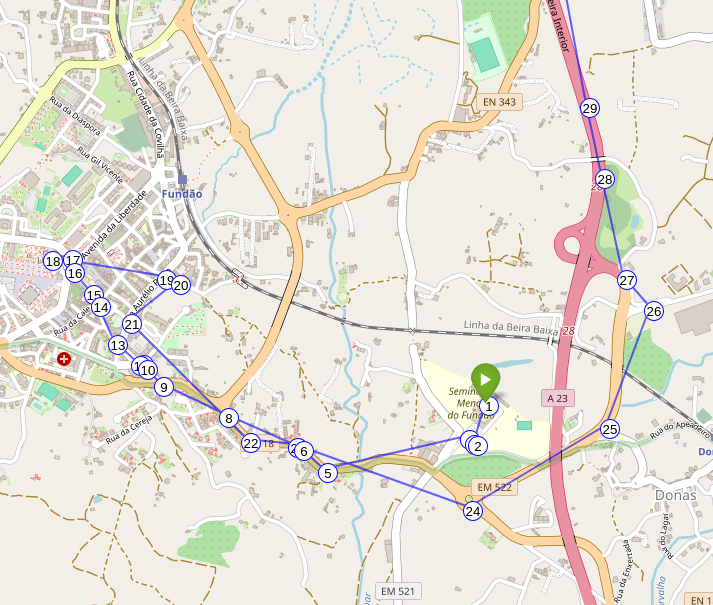}
        \caption[Detected backtrack in route from Farm to Factory]{Detected backtrack in route from Farm to Factory. Notice how the truck went first to the left instead of directly to its destination. Then after a small detour in the city it returns to its original route.}
    \label{fig:invalid-backtrack}
\end{figure}

The \textit{backtrack rule} ensures that the trucks consistently move toward their destination, detecting detours even if the vehicle remains within allowed areas. We evaluated two shipments: one following a correct route and one with an intentional detours. \emph{ChainGuards} flags the points where the route moves away from the destination. Figure \ref{fig:valid-geofence} shows a shipment that passes verification, while Figure \ref{fig:invalid-backtrack} depicts a detouring shipment that triggers alerts on the deviating segments.

The \textit{threshold rule} ensures that products remain within safe environmental and logistical limits. \emph{ChainGuards} verifies both product conditions (e.g., temperature, humidity) and process timings (e.g., shipment duration, hand-over intervals).
For the temperature, alerts are raised when readings exceed predefined limits. However, to distinguish minor or brief violations from significant ones, a cumulative severity method considers both the magnitude and duration of threshold breaches. Severity is calculated by summing the average excess above the limit across consecutive readings, weighted by the time between them (Equation~\ref{eq:1}).

\begin{equation}
\label{eq:1}
\text{Severity} = \sum_{i=1}^{n-1} \frac{(E_i + E_{i+1})}{2} \cdot \Delta t_i
\end{equation}

\indent where \(E_i = \max(0, T_i - T_\text{max})\) represents the excess of the \(i\)-th reading above the threshold \(T_\text{max}\), and \(\Delta t_i\) is the time in minutes between consecutive readings.

In this study, only maximum temperatures (\(T_\text{max} = 4^\circ\text{C}\) sampled every 10 minutes) were evaluated, as highlighted by the expert (c.f. Section~\ref{sec:system-overview}). Table~\ref{tab:table-temperature} shows that the simple threshold flags three alerts, while the cumulative method only triggers alerts for sustained exceedance. Threshold rules also apply to logistics. The shipment timeout rule flags deliveries exceeding expected durations, while hand-over verification monitors delays between supply chain transfers. For example, the shipment in Figure \ref{fig:valid-geofence} meets timing requirements, whereas the detouring shipment in Figure \ref{fig:invalid-backtrack} is likely to fail.



\begin{table}[ht]
\centering
\footnotesize
\caption[Threshold and cumulative verification of Temperature]{Threshold and cumulative verification of Temperature. Threshold = 4ºC. Cumulative threshold = 30.}
\label{tab:table-temperature}
\begin{tabular}{ >{\raggedleft\arraybackslash}m{2.5cm} |
                >{\raggedright\arraybackslash}m{2.3cm} |
                >{\raggedleft\arraybackslash}m{4.0cm} |
                >{\raggedright\arraybackslash}m{1.5cm} } 
 \hline
 \textbf{Temperature (ºC)} & \textbf{Fixed Threshold} & \textbf{Cumulative (excess x time)} & \textbf{Cumulative Outcome}\\ [1ex] 
 \hline
 3.2  & OK    & 0  & OK\\
 4.5  & Alert & 0.5(ºC) × 10(min) = 5  & OK\\
 5.0  & Alert & 1.0(ºC) × 10(min) = 10 & OK\\
 3.8  & OK    & 15 & OK\\
 3.5  & OK    & 15 & OK\\
 2.9  & OK    & 15 & OK\\
 3.1  & OK    & 15 & OK\\
 6.0  & Alert & 2.0(ºC) × 10(min) = 20 & Alert \\
 3.0  & OK    & 35 & ---\\
 \hline
\end{tabular}
\end{table}

Previous examples relied on a single sensor, which is susceptible to false positives or negatives in case of malfunction. Using additional devices enables a multiple source of data (MSoD) model, in which readings can be correlated to improve reliability. 
Policies may specify whether verification follows SSoD or MSoD, with a Mahalanobis-weighted Kalman filter 
to merge data and mitigate noise and outliers.

\begin{figure}[htp]
    \centering
    \begin{minipage}[t]{0.225\textwidth}
        \centering
        \includegraphics[width=\linewidth]{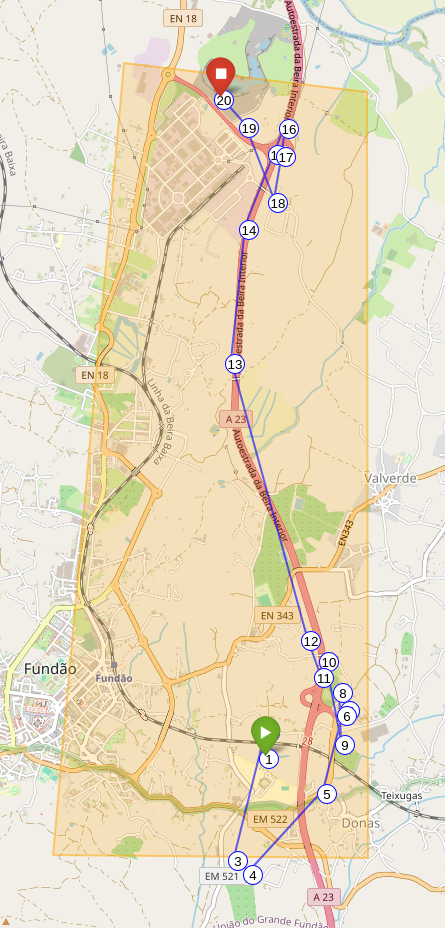}
        \caption{Route captured by the second device}
        \label{fig:second-device}
    \end{minipage}
    \begin{minipage}[t]{0.225\textwidth}
        \centering
        \includegraphics[width=\linewidth]{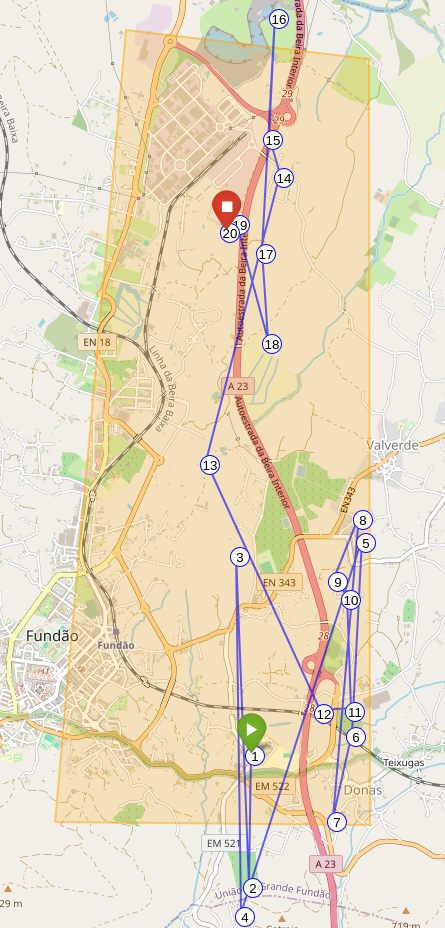}
        \caption{Route captured by the third device}
        \label{fig:third-device}
    \end{minipage}\hfill    
\end{figure}

\begin{figure}[htp]
    \centering
    \begin{minipage}[t]{0.225\textwidth}
        \centering
        \includegraphics[width=\linewidth]{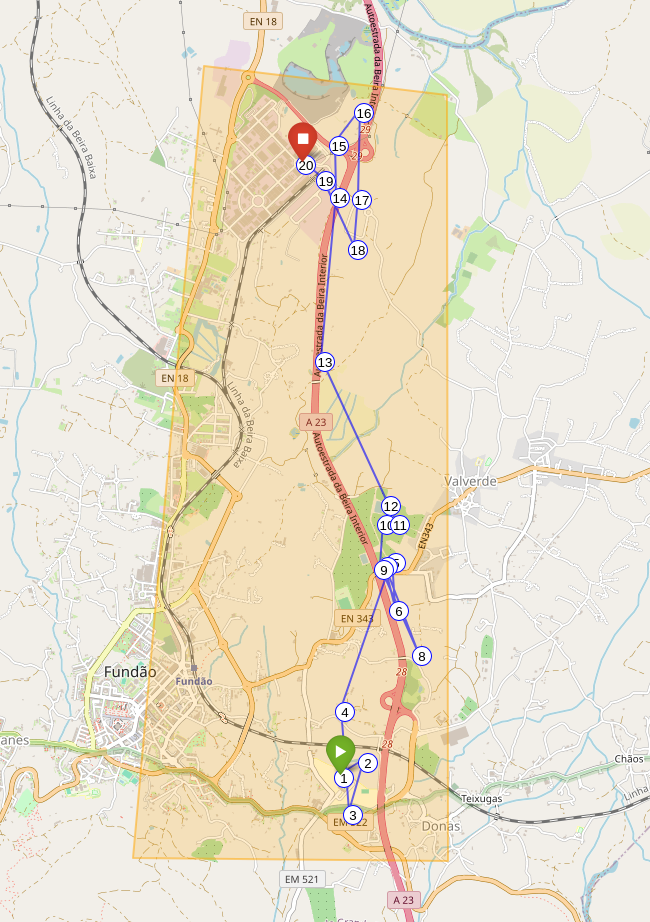}
        \caption{Merged route with Weighted-KF}
        \label{fig:merged-msod}
    \end{minipage}
    \begin{minipage}[t]{0.225\textwidth}
        \centering
        \includegraphics[width=\linewidth]{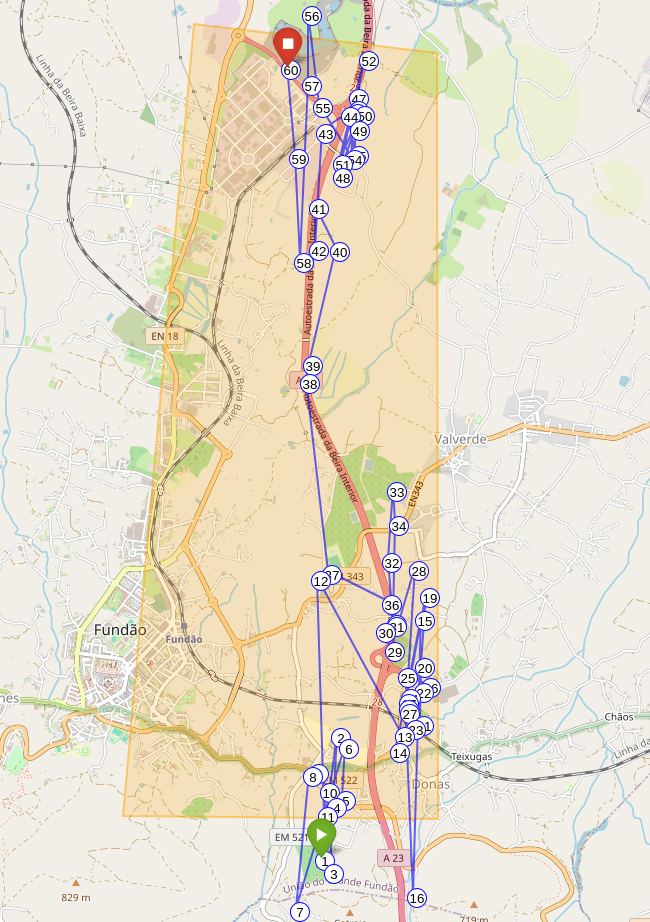}
        \caption{Merged route with normal KF}
        \label{fig:merged-ssod-map}
    \end{minipage}\hfill    
\end{figure}

We simulated three GPS devices along the route in Figure~\ref{fig:valid-geofence}: the first was accurate, the second had moderate noise (Figure~\ref{fig:second-device}), and the third had significant noise and outliers (Figure~\ref{fig:third-device}). After applying the algorithm, the merged path provides a smoother and more reliable reconstruction (Figure~\ref{fig:merged-msod}).

Even though two devices individually fail verification, the filtered consensus ensures the rule passes. By contrast, merging all devices equally with a simple Kalman filter smooths the data but fails verification (Figure~\ref{fig:merged-ssod-map}).

\emph{ChainGuards} supports cross-verification to ensure consistency between claims made by multiple stakeholders about the same item. A claim represents a stakeholder’s reported data or observation, such as sensor readings from trucks owned by different companies.
In a DSoD setting, each claim is verified independently and outcomes are compared. Discrepancies between results highlight potential issues requiring further investigation. Beyond rule outcomes, the system can check journey-wide attributes (e.g., parent identifiers, weight, etc.) for consistency.

Table~\ref{tab:journey-discrepancy} shows claims made by three stakeholders for a product. Variety and color pass without issues, while weight and parentage discrepancies are detected. 
All alerts provide details of the detected discrepancies.

\begin{table}[ht]
\centering
\footnotesize
\caption{Journey attribute consistency across stakeholders}
\label{tab:journey-discrepancy}
\begin{tabular}{ >{\raggedright\arraybackslash}m{1.8cm} 
                 | >{\raggedright\arraybackslash}m{2cm} 
                 | >{\raggedright\arraybackslash}m{2cm}
                 | >{\raggedright\arraybackslash}m{2cm}
                 | >{\raggedright\arraybackslash}m{2cm} } 
\hline
\textbf{Stakeholder} & \textbf{Parent ID} & \textbf{Weight (kg)} & \textbf{Variety} & \textbf{Color Grade} \\ \hline

Producer company     & Lot-123 & 1000 & Cherry-A & Red \\ \hline
Sensor company       & Lot-122 & 998  & Cherry-A & Red \\ \hline
Receiver company     & Lot-123 & 950  & Cherry-A & Red \\ 
\hline
\end{tabular}
\end{table}

\subsection{Pilot} 
\label{sec:pilot}

The pilot was conducted in the Fundão region of Portugal, where PGI (Protected Geographical Indication) certification\footnote{A certification granted by the European Union indicating that a product possesses qualities, reputation, or characteristics linked to a specific geographical region.} enforces strict standards for cherry size, color and origin. The supply chain involved CMF farm (production), CF factory (sorting and packaging), and retailer SN (sales), with coordinated transportation. The flow of the product throughout the supply chain between each stakeholder is represented in Figure \ref{fig:usecase-supplychain}.

At the farm, production was documented digitally, and cherries were collected into unmarked boxes, limiting direct traceability. At the factory, boxes were assigned batch identifiers, sorted by quality, 
and assembled into pallets with Serial Shipping Container Codes (SSCC). Retailers scanned and disassembled pallets to fulfill store orders, generating new SSCC codes, but individual boxes 
remained untraceable at the consumer level. To address these gaps, the pilot deployed sensors: Tags on boxes and pallets, Mobile Gateways with GPS, Fixed Gateways at key facilities, a mobile app for managing Tag associations and a LoRa network for long-range connectivity. Although full end-to-end traceability was limited by partial data sharing, inconsistent formats and sensor issues, a coherent trace was reconstructed by combining available data segments, 
enabling verification and evaluation of \emph{ChainGuards} under practical conditions.

\begin{figure}[htp]
    \centering
    \includegraphics[scale=0.35]{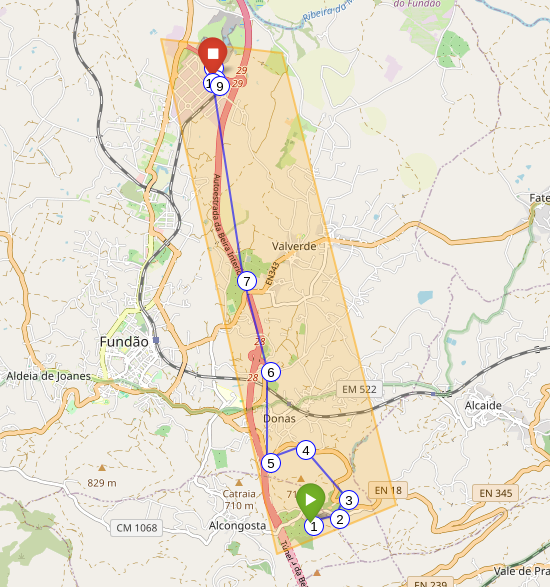}
    \caption{Route collected from the farm in Alcongosta}
    \label{fig:alcongosta-farm}    
\end{figure}

Before starting the pilot we defined a verification policy to evaluate the trace and detect deviations from expected behavior. For the first phase, we selected a complete route from farm to factory, collected on July 7th from a farm in Alcongosta. 
Figure~\ref{fig:alcongosta-farm} shows the recorded route and geofence.

The shipment (lot ALC250000000545, tracked with Tag R1) passed all checks: geofence, backtrack, radius and shipment timeout. Shipping lasted $\sim$16 minutes, consistent with expectations. On the same day, the factory registered multiple lots arrivals (CER250000000554 and CER250000000545). Because factory identifiers are generated after weighing, the sensor company issued mapping events to align field and factory identifiers. This mapping linked the field lot (ALC250000000545) to the factory lot (CER250000000545) and subsequently to pallet 356021630000259170, with a new Tag (T7) assigned to capture later shipping events. Discrepancy detection compared the sensor company mapping with the factory’s internal system mapping for the same pallet. Two discrepancies were identified:

\begin{itemize}
    \item \textbf{Variety}: Factory recorded ``T'' (likely an internal code), sensor company recorded ``Lapins.''
    \item \textbf{Lineage (Parent identifiers)}: Conflicting identifiers highlighted a true inconsistency.
\end{itemize}

\begin{figure}[htp]
    \centering
        \includegraphics[width=0.4\linewidth]{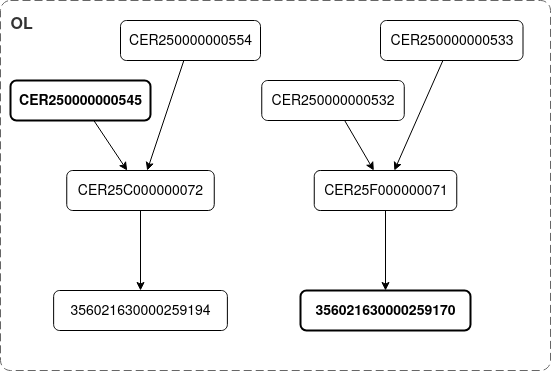}
        \caption{DAG representing factory lineage claim}
        \label{fig:ol-claim}  
\end{figure}

\begin{figure}[htp]
    \centering
        \centering
        \includegraphics[scale=0.4]{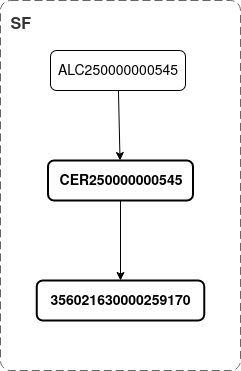}
        \caption{DAG representing sensor company lineage claim}
        \label{fig:sf-claim}
\end{figure}

Figures \ref{fig:ol-claim} and \ref{fig:sf-claim} illustrate these discrepancies as graphs, 
highlighting the pallet and diverging parent identifiers. The pallet was then shipped to a depot in Porto on July 9th. From this point, only the sensor company provided data (SSoD), preventing cross-verification. Continuous GPS tracking was unavailable due to the refrigerated truck with metallic structure, so only departure and arrival events were recorded via mobile app.

\begin{figure}[htp]
    \centering
        \centering
        \includegraphics[scale=0.2825]{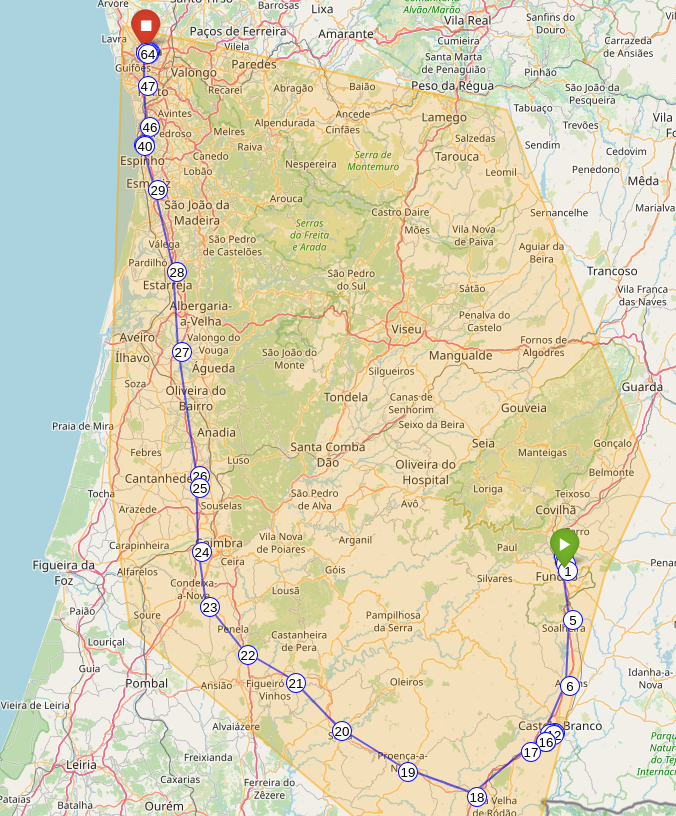}
        \caption{Route from factory in Fundão to depot in Porto}
        \label{fig:depot-route}
\end{figure}

To illustrate the route, we referenced a similar shipment from August 8th, 2024 (Figure~\ref{fig:depot-route}). This route respected the geofence, though backtrack verification flagged anomalies at the start (sensor noise), near Castelo Branco (stop), at a small logistics facility (overnight stay) and at the destination (sensor noise). Noisy readings at start and end caused radius verification to fail.

Using departure and arrival events from July 9th, the calculated shipping time was $\sim$1 hour 45 minutes. This failed the shipment timeout rule, not because the shipment was delayed, but because the recorded duration is unrealistically short, suggesting that the departure event was logged later than the actual departure.

The pallet was later transformed and shipped to a retailer, but no location data was available, making further verification impossible. The shipment duration could be assessed in principle, but without the retailer’s location, thresholds could not be defined. The pilot evaluation shows that finding the anomalies helps to identify where data does not align as expected. Noticing these differences creates opportunities to understand their causes and make adjustments to the processes. With time, gradual improvements can lead to better quality.

\subsection{Performance Analysis}

Performance was evaluated on a Hyperledger Fabric v2.5.9 network deployed on a laptop (Intel i7-9750H, 32\,GB RAM, NVMe SSD). We measured (i) ingestion latency, storing parsed system objects versus raw events, and (ii) verification latency for varying batch sizes. For each transaction we recorded start/end times and computed average, minimum, maximum, and standard deviation.

\begin{table}[ht]
\centering
\footnotesize 
\caption{Comparison of Transaction Latencies: Baseline vs Chainguards Ingestion (1000 events)}
\begin{tabular}{  >{\raggedright\arraybackslash}m{3cm} 
                 | >{\raggedleft\arraybackslash}m{2cm}
                 | >{\raggedleft\arraybackslash}m{2cm}
                 | >{\raggedleft\arraybackslash}m{2cm}
                 | >{\raggedleft\arraybackslash}m{2cm}  } 
 \hline
\textbf{Type} & \textbf{Min (ms)} & \textbf{Max (ms)} & \textbf{Avg (ms)} & \textbf{Std Dev (ms)} \\
\hline
Baseline & 2140 & 2193 & 2152.56 & 5.67 \\
Chainguards Ingest & 2138 & 2223 & 2153.26 & 6.29 \\
\hline
\end{tabular}
\label{tab:baseline_vs_ingest}
\end{table}

Table~\ref{tab:baseline_vs_ingest} shows that ingestion overhead is negligible: for 1000 sequential events the additional cost is $\sim$0.03\% compared to raw storage. Table~\ref{tab:verify_latency} reports verification latency for 10, 100, and 1000 reports. Storing sensor readings as individual Points increases verification time for 1000 reports by $\sim$100\,ms, but enables parallel ingestion, reducing total ingestion time from $\sim$33 minutes (sequential) to under 1 minute (parallel). MSoD verification with three devices and $\sim$1050 events achieved a similar latency (2358\,ms), confirming scalability with minor overhead. In a two-organization Raft/TLS setup, absolute latencies increased slightly ($\sim$1\%), while relative overhead remained stable, indicating that \emph{ChainGuards} does not become a bottleneck as network complexity increases.

\begin{table}[ht]
\centering
\footnotesize 
\caption{Verification Latency for Different Event Batch Sizes}
\begin{tabular}{  >{\raggedright\arraybackslash}m{2cm} 
                 | >{\raggedleft\arraybackslash}m{2cm}
                 | >{\raggedleft\arraybackslash}m{2cm}
                 | >{\raggedleft\arraybackslash}m{2cm}
                 | >{\raggedleft\arraybackslash}m{2cm}  } 
\hline
\textbf{Batch Size} & \textbf{Min (ms)} & \textbf{Max (ms)} & \textbf{Avg (ms)} & \textbf{Std Dev (ms)} \\
\hline
10 events & 2154 & 2170 & 2160.90 & 4.86 \\
100 events & 2172 & 2189 & 2179.70 & 5.52 \\
1000 events & 2330 & 2366 & 2346.00 & 12.47 \\
\hline
\end{tabular}
\label{tab:verify_latency}
\end{table}

\section{Discussion}
\label{sec:discussion}

The evaluation shows that \emph{ChainGuards} can reliably verify supply-chain data while introducing only minimal performance overhead.

In simulation-based experiments, verification rules such as geofencing, backtrack detection, and shipment timeouts effectively enforced route compliance and shipment integrity, to confirm that products remained within expected regions and progressed toward their intended destinations. Environmental monitoring rules successfully detected both transient sensor anomalies and sustained deviations from acceptable conditions. Moreover, multi-sensor aggregation and discrepancy detection mechanisms improved data reliability across stakeholders by identifying inconsistent or conflicting reports (Table~\ref{tab:system-requirements}).

The Fundão cherry pilot study validated many of these mechanisms under real-world operational conditions. \emph{ChainGuards} consistently verified shipment routes (R1, R2), detected discrepancies between records reported by the sensor service provider and the processing facility (R8), and supported backward traceability across the supply chain. Quality testing and recall support were partially implemented (R4, R5), enabling limited integration of laboratory results with provenance data. However, full automation of these processes, particularly integration with stock and inventory management systems, remained outside the scope of the prototype. Alerts and notifications were correctly generated in response to detected deviations and inconsistencies (R7), facilitating timely human intervention and reinforcing accountability among participating organizations.

From a performance perspective, \emph{ChainGuards} remained robust even in more complex blockchain configurations involving multiple peers and Raft-based consensus. Advanced verification mechanisms, including multi-sensor and multi-stakeholder checks (R6, R8), introduced only marginal additional latency. These results confirm that the system scales efficiently as verification complexity and network size increase, without compromising responsiveness.

\emph{ChainGuards} satisfies the majority of expert-derived requirements (Table~\ref{tab:system-requirements}) by effectively verifying product origin, route integrity, and inter-organizational data consistency, while providing actionable alerts and strong traceability guarantees. The remaining limitations are mainly due to incomplete or unavailable operational data, rather than the system architecture or implementation themselves. This suggests that improvements in data availability, sensor coverage, and system integration could further improve the framework’s effectiveness in real-world supply chains.

\section{Conclusion}
\label{sec:conclusion}

\emph{ChainGuards} bridges the gap between raw sensor outputs and trustworthy supply-chain records by introducing a decentralized, rule-driven verification layer. Implemented as chaincode on a Hyperledger Fabric network, the system ingests events from each business phase and applies context-aware, phase-specific verification rules before data is permanently recorded. \emph{ChainGuards} supports three complementary verification models: single source (SSoD), multiple sources (MSoD), and diverse sources of data (DSoD); and employs a structured object model that links product journeys and supply-chain steps to sensor-derived evidence.

The system was evaluated using both simulated data and real-world datasets.
Simulations were used to generate controlled scenarios with injected anomalies, while real-world pilot data provided practical validation under realistic operating conditions.
End-to-end experiments show how \emph{ChainGuards} can reliably detect out-of-range sensor readings, geofence violations, route deviations, hand-over delays, and other integrity breaches, while maintaining low performance overhead. By enabling automated, policy-driven verification across organizational boundaries, \emph{ChainGuards} reduces reliance on manual audits and provides a foundation for automated and transparent supply-chain monitoring.

Several directions are open for future research and development. First, the requirements-gathering process can be extended by involving a broader and more diverse set of domain experts, including logistics operators, regulators, and auditors, to further refine verification rules and governance mechanisms. Second, greater flexibility in handling detour-related alarms will be explored, enabling the system to better distinguish between legitimate operational deviations (e.g., traffic or safety-related rerouting) and suspicious behavior. Additionally, the evaluation will be extended to larger deployments, longer timeframes, and comparisons with alternative verification approaches.

\section*{Acknowledgements}

Work supported by national funds through Fundação para a Ciência e a Tecnologia, I.P. (FCT) under projects UID/50021/2025 (DOI: \url{https://doi.org/10.54499/UID/50021/2025}) and UID/PRR/50021/2025 (DOI: \url{https://doi.org/10.54499/UID/PRR/50021/2025}) and by Blockchain.PT – Decentralize Portugal with Blockchain Agenda, (Project no 51), WP 1: Agriculture and Agri-food, Call no 02/C05-i01.01/2022, funded by the Portuguese Recovery and Resilience Program (PRR), The Portuguese Republic and The European Union (EU) under the framework of Next Generation EU Program.

\clearpage

\bibliographystyle{unsrtnat}
\bibliography{references}

\end{document}